\documentclass{ws-procs9x6}

\begin{document}

\title{MiniBooNE Status}

\author{J.~L. RAAF\footnote{\uppercase{T}his work is partially supported by grant \uppercase{PHY}-0244915 of the \uppercase{N}ational \uppercase{S}cience \uppercase{F}oundation.}~~for the BooNE collaboration\footnote{\uppercase{S}ee http://www.boone-fnal.gov/cgi-bin/collaboration/ for entire collaboration list.}}

\address{University of Cincinnati \\
  Department of Physics, M.L. 0011\\
  Cincinnati, OH 45221, USA\\
  E-mail: jlraaf@fnal.gov}

\maketitle

\abstracts{MiniBooNE, the Mini Booster Neutrino Experiment at Fermilab, will confirm or refute the existence of the neutrino oscillation signal seen by the Liquid Scintillator Neutrino Detector (LSND) Experiment at Los Alamos National Laboratory. The experiment will search for the appearance of electron neutrinos in a beam of muon neutrinos. This work presents preliminary results from the first round of analyses of charged current quasi-elastic events, neutral current $\pi^{0}$ events, and neutral current elastic events. The neutrino oscillation analysis is not presented in this work; it is a blind analysis which will not be presented until the full set of data has been collected.}

\section{Introduction}
\label{sec:intro}
MiniBooNE, the Mini Booster Neutrino Experiment at Fermilab, has been running smoothly and taking data for over a year now. The experiment intends to make a definitive statement about the neutrino oscillation signal seen by the Liquid Scintillator Neutrino Detector (LSND) Experiment at Los Alamos National Laboratory \cite{LSND}. Since both solar and atmospheric neutrino oscillations have been well-established in recent years, confirmation of the existence of another oscillation signal with its own distinct $m^{2}$ splitting would indicate a need for further extensions to the Standard Model in order to accommodate this signal.

The Fermilab Booster operates on a 15 Hz cycle, sending between 2 and 3 Hz of its total to MiniBooNE. This 8 GeV proton beam from the Booster is directed into a 71 cm long beryllium target located in a magnetic focusing horn. The horn sends positively charged particles into a 50 meter decay region. Any charged particles that do not decay in the 50 meters are absorbed by or decay in the absorber located at the downstream end of the decay region. The neutrinos produced in the charged particle decays then travel through approximately 500 meters of dirt to reach the detector. By the time the beam reaches the detector it is very pure -- almost entirely made up of muon neutrinos. Only about 0.5\% of the beam is electron neutrinos. The average energy of a neutrino entering the MiniBooNE detector is 800 MeV. 

The detector consists of a sphere within a sphere, where the inner signal region is optically isolated from the outer veto shell \cite{BooNEproposal}. The entire detector is instrumented with 8-inch Hamamatsu photomultiplier tubes (PMTs): 1280 in the main signal region of the detector (10\% coverage) and 240 in the veto region \cite{PMTtest}. The entire detector is filled with approximately 800 tons of pure mineral oil \cite{Oiltest}.

\begin{figure}[ht]
\hfill
\begin{minipage}{1.4in}
\epsfxsize=1.4in\epsfbox{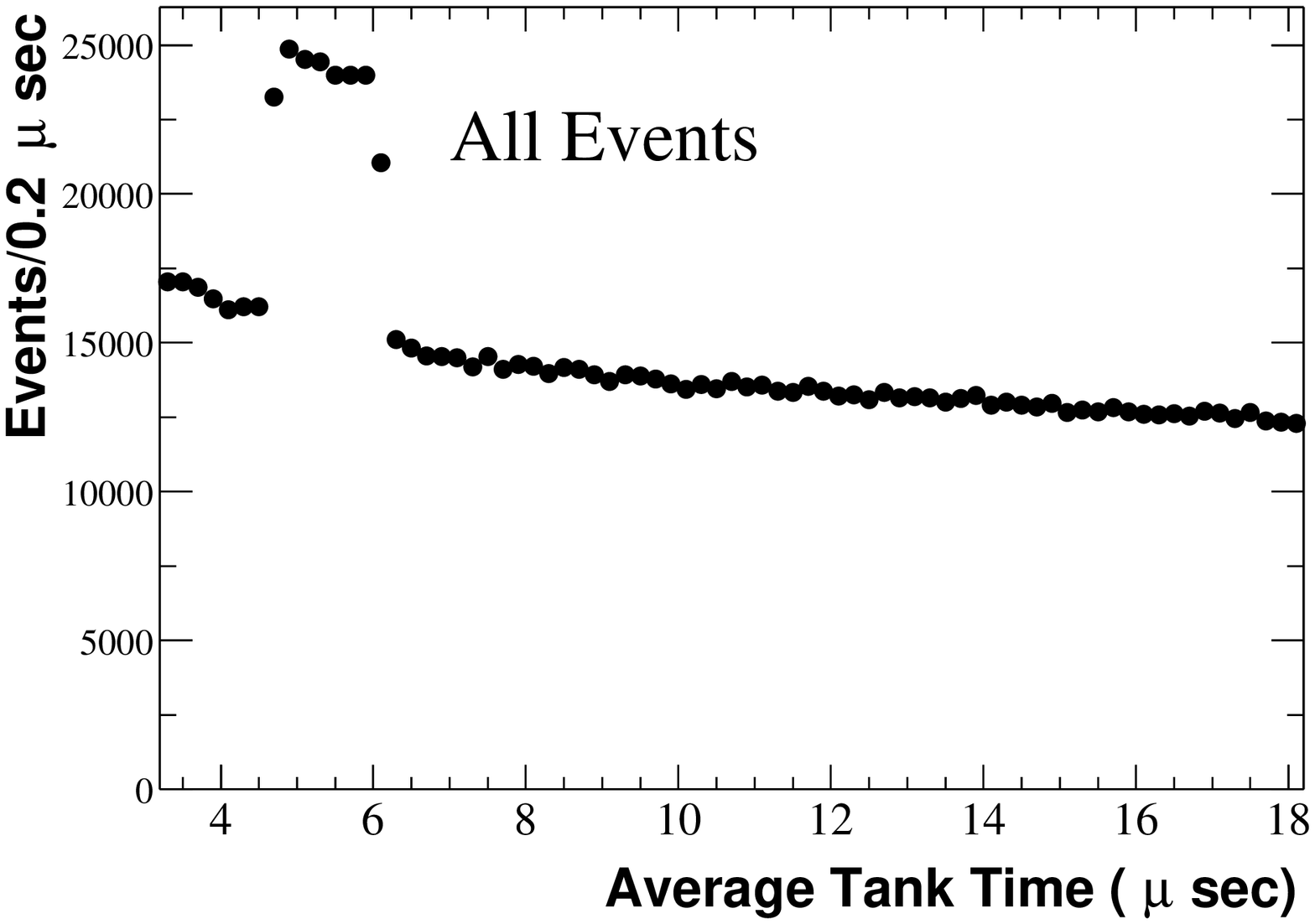}
\end{minipage}
\hfill
\begin{minipage}{1.4in}
\epsfxsize=1.4in\epsfbox{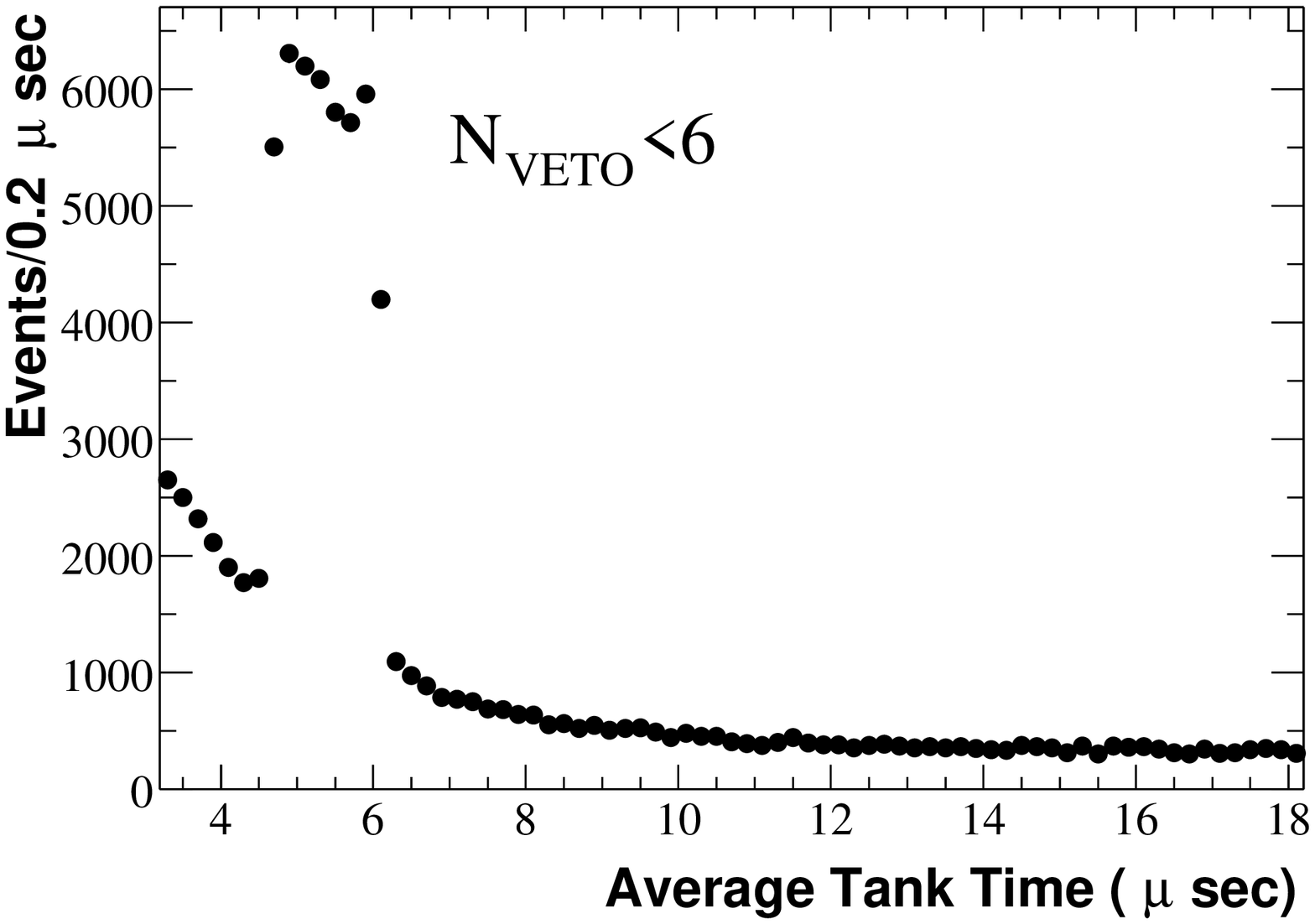}
\end{minipage}
\hfill
\begin{minipage}{1.4in}
\epsfxsize=1.4in\epsfbox{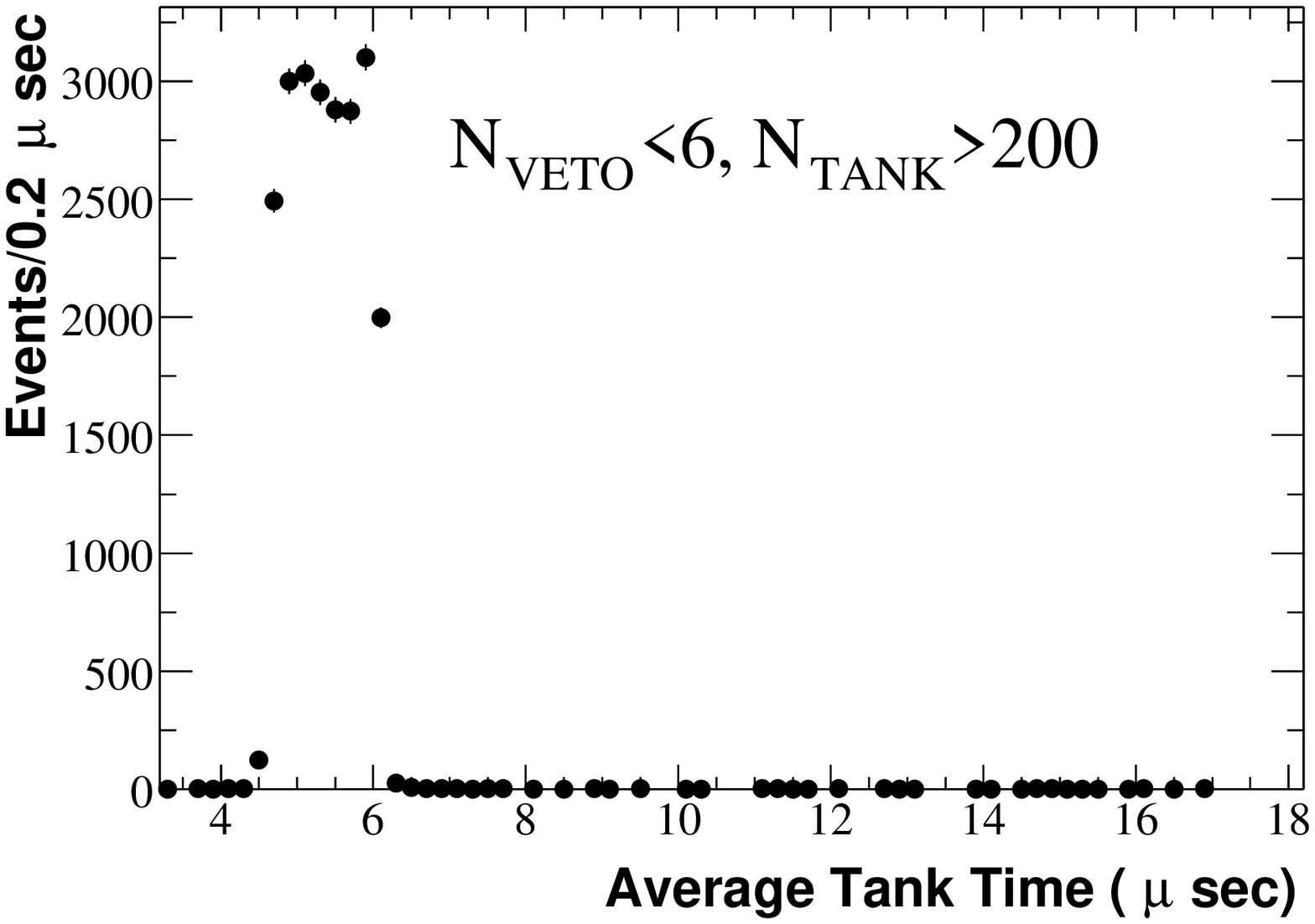}
\end{minipage}
%\hfill\mbox{}
\caption{The left panel shows recorded tank activity during $19.2~\mu \mathrm{s}$ beam window with no cuts. The center panel adds a simple veto cut. The right panel adds a further requirement of greater than 200 hits in the main volume of the detector. 
\label{beamtiming}}
\end{figure}

Beam arrives at the target in $1.6~\mu \mathrm{s}$ bursts. The data acquisition system (DAQ) receives a signal from the Booster indicating that beam will be arriving imminently. This signal triggers the DAQ to open a window and record all detector activity for $19.2~\mu \mathrm{s}$, starting $4.8~\mu \mathrm{s}$ before beam arrives. Fig.~\ref{beamtiming} shows the distribution of events during the beam window. The left panel shows the $1.6~\mu \mathrm{s}$ wide peak of the beam particles arriving in the middle of the beam window.

With the application of only two very simple cuts, the background on the distribution can be drastically reduced. Requiring no more than 6 hits in the veto region removes events with a coincident cosmic muon occurring in the beam window, shown in the center panel of Fig.~\ref{beamtiming}. The further requirement of at least 200 hits in the main tank region, shown in the right panel, largely removes the chance of an electron from muon decay occurring during the beam window. With only these two cuts, the signal to background ratio is greater than 1000:1. As further confirmation that the particles observed are actually coming from the beam, we check the reconstructed direction of these particles with respect to the beam direction. The majority of the particles align with the beam direction.

\section{Neutrino Interaction Analyses}
The three main analyses that are currently underway at MiniBooNE are charged current quasi-elastic scattering (CCQE), neutral current $\pi^{0}$ production (NC $\pi^{0}$), and neutral current elastic scattering (NCE). At MiniBooNE energies, about 39\% of neutrino interactions are due to CCQE, 7\% NC $\pi^{0}$, and 17\% NCE. From the dataset of about $1\times10^{20}$ protons on target analyzed, we already have an abundance of events to study for each of these analyses.

\subsection{Charged Current Quasi-Elastic Events}

\begin{figure}[ht]
\begin{minipage}{2.1in}
\epsfxsize=2.1in\epsfbox{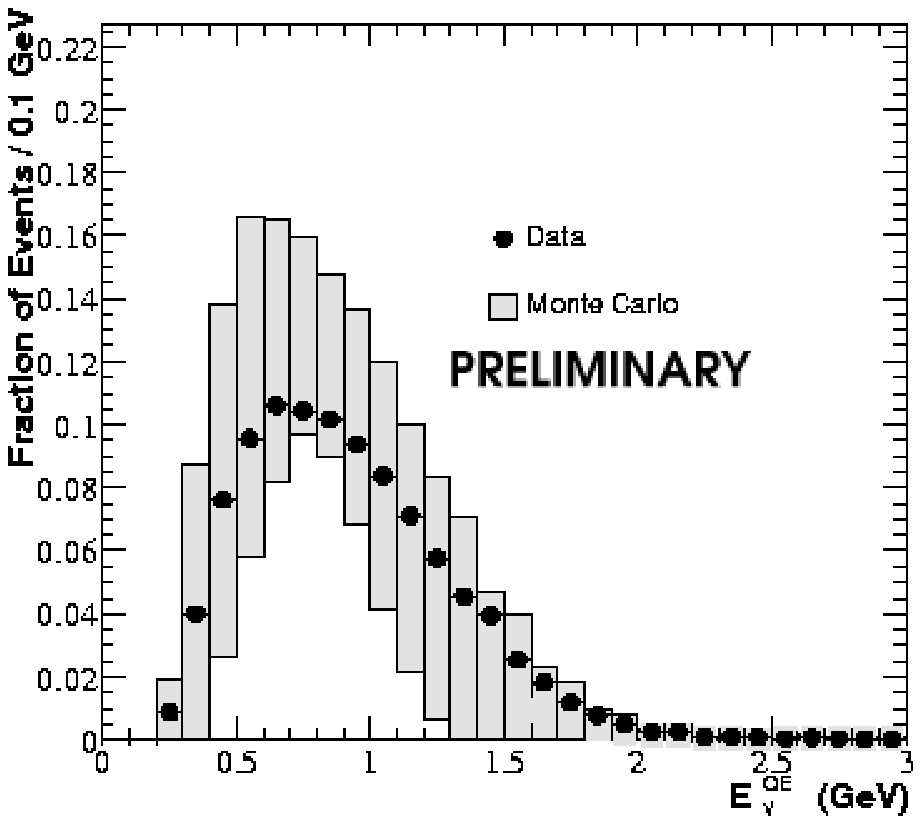}
\end{minipage}
\hfill
\begin{minipage}{2.1in}
\epsfxsize=2.1in\epsfbox{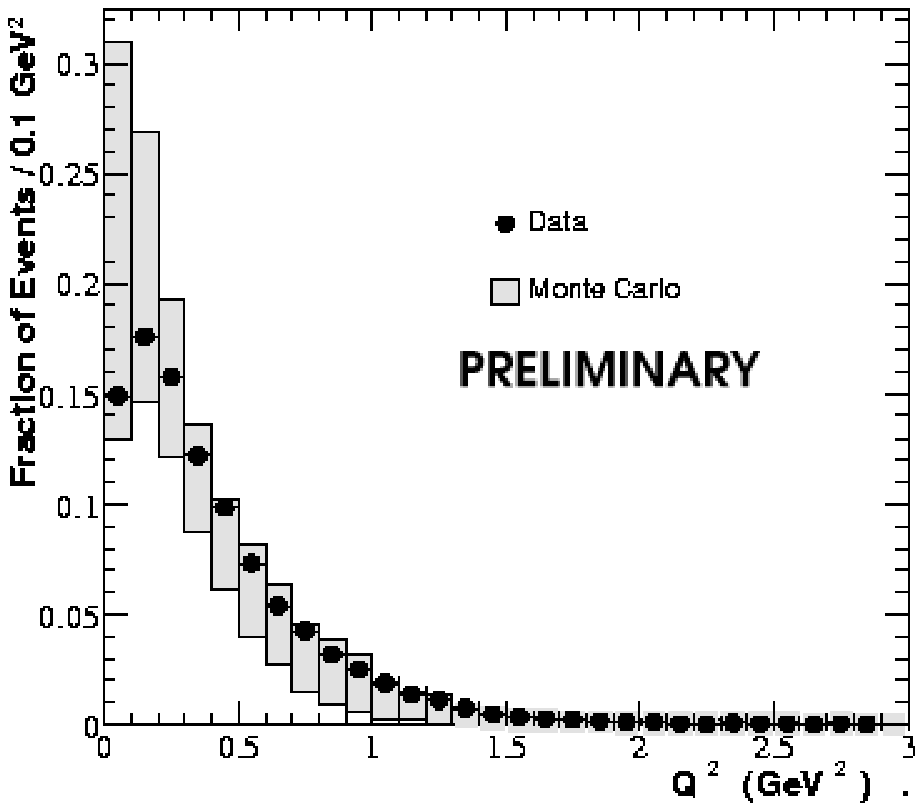}
\end{minipage}
\caption{Reconstructed neutrino energy, $E_{\nu}^{QE}$, (left) and negative square of the four-momentum transfer, $Q^2$, between the neutrino and the muon (right) for $\nu_{\mu}$ charged current quasi-elastic events. The black dots indicate data, the grey rectangles indicate the Monte Carlo predictions including the current level of systematic uncertainties. All distributions are normalized to unit area.
\label{CCQE}}
\end{figure}

Since charged current quasi-elastic interactions are among the most prevalent events at MiniBooNE energies, there are already more than 30,000 events in the study sample. It is important to understand these interactions because $\nu_{e}$ CCQE events with similar kinematics and cross section will be the main signal for the $\nu_{\mu} \rightarrow \nu_{e}$ oscillation search, and these interactions will help to constrain the intrinsic $\nu_{e}$ flux in the beam. In addition, comparison of the actual CCQE cross section with the expected cross section will allow MiniBooNE to make a measurement of $\nu_\mu$ disappearance in this channel for $\Delta\mathrm{m}^{2} \sim 0.1-10~\mathrm{eV}^{2}$. The CCQE event selection requires a series of cuts to obtain a high-purity sample of well-reconstructed $\nu_{\mu}$ events. The cosmic ray-induced activity is reduced by the method discussed in section \ref{sec:intro}. Events in which the neutrino energy cannot be reliably reconstructed are rejected via a fiducial volume cut. Finally, the selection requires a single muon-like \v{C}erenkov ring and scintillation light production consistent with a  $\nu_{\mu}n\rightarrow \mu^{-}p$ interaction.

The CCQE ~analysis \cite{jocelyn} determines the neutrino energy, $E_{\nu}$, using the outgoing muon energy, $E_{\mu}$, and outgoing muon angle relative to the neutrino beam direction, $\cos \theta_{\mu}$. Reconstructed neutrino energy for data is compared to Monte Carlo predictions, both normalized to unit area, in the left panel of Fig.~\ref{CCQE}. The error bars on this figure include our current assessment of the effect of the major sources of systematic uncertainties. In particular, systematic errors associated with the $\nu_{\mu}$ flux, $\nu_{\mu}$ CCQE cross section, and the properties of light production and transmission in the MiniBooNE detector are taken into account. Substantial reductions in these systematics are expected as the analysis progresses.

The right panel of Fig.~\ref{CCQE} ~shows the measured and expected distributions in $\mathrm{Q}^2$, the negative of the squared four-momentum transfer between the incoming neutrino and the outgoing muon in the two-body CCQE scattering process. The behavior of the $\mathrm{Q}^2$ distribution at low $\mathrm{Q}^2$ values is particularly important in experimentally validating how well nuclear effects are modeled in the Monte Carlo description of neutrino interactions on the bound target nucleons of the MiniBooNE detector.

\subsection{Neutral Current $\pi^{0}$ Events}

\begin{figure}[ht]
\centerline{\epsfxsize=2.0in\epsfbox{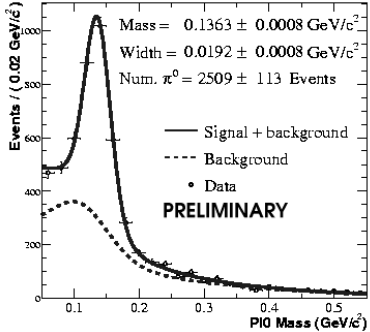}}
\caption{\label{fig:pi0_mass}Reconstructed invariant mass of beam triggers (open circles with statistical error bars). Fitted shapes are Monte Carlo-based parameterizations of the contribution from background (dashed curve) and signal (NC resonant and coherent $\pi^{0}$'s, solid minus dashed curve). Note that some of the background events contain $\pi^{0}$'s, so the peak near $m_{\pi^{0}}$ in the dashed curve is expected.}
\end{figure}

\begin{figure}[ht]
\hfill
\begin{minipage}{1.4in}
\epsfxsize=1.4in\epsfbox{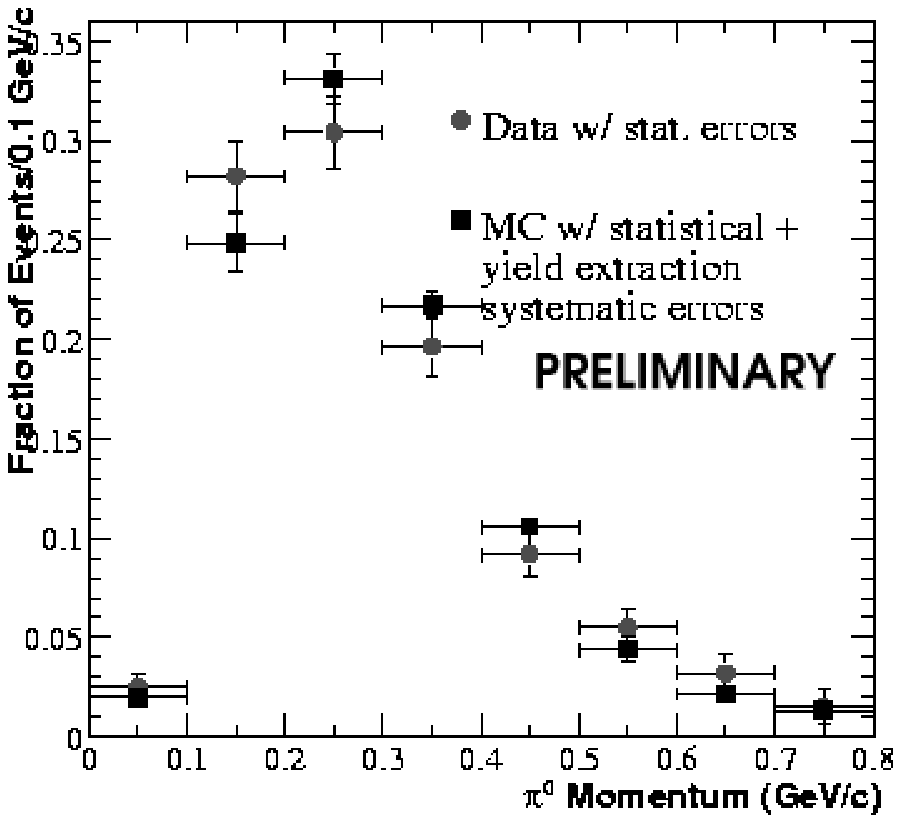}
\end{minipage}
\hfill
\begin{minipage}{1.4in}
\epsfxsize=1.4in\epsfbox{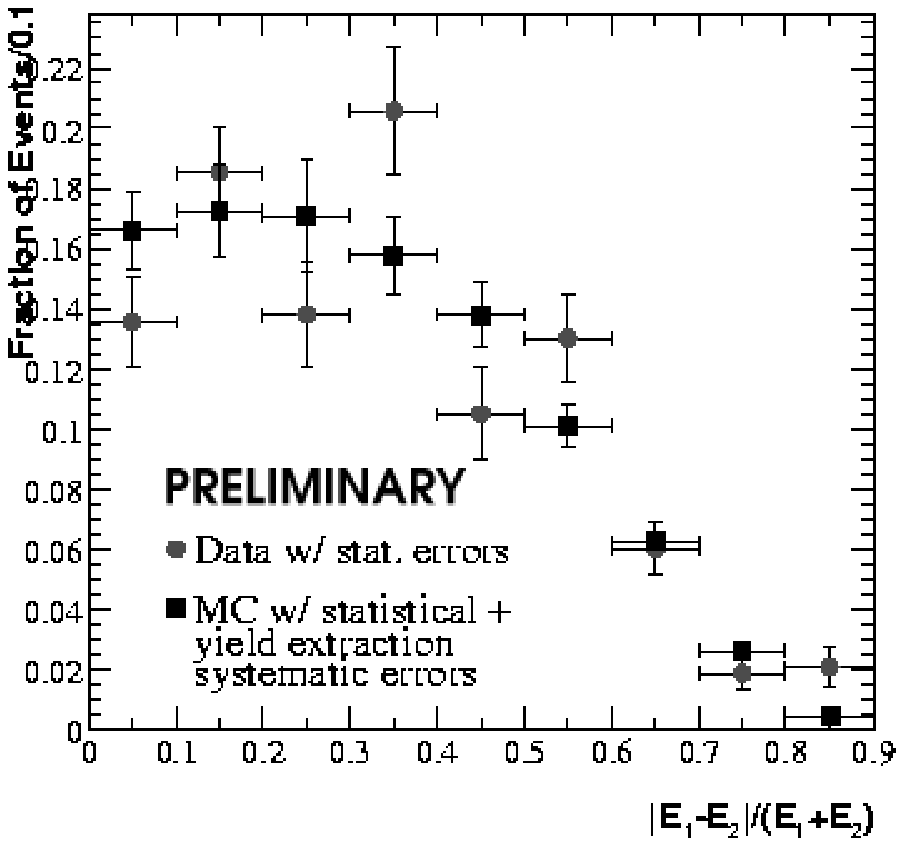}
\end{minipage}
\hfill
\begin{minipage}{1.4in}
\epsfxsize=1.4in\epsfbox{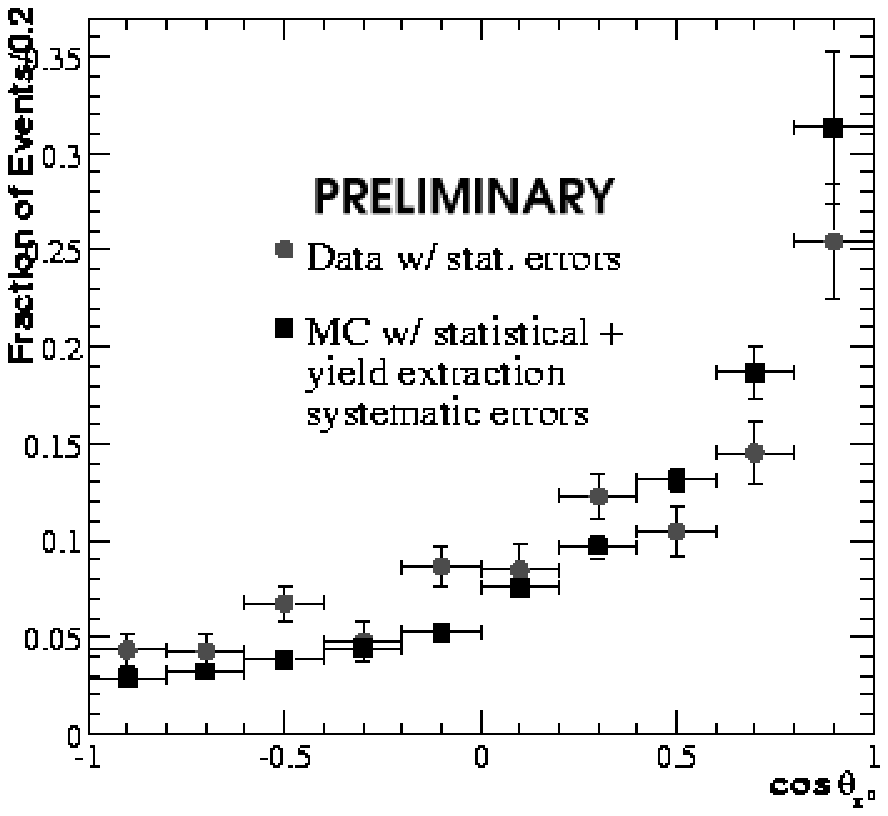}
\end{minipage}
\caption{Extracted $\pi^{0}$ yields in $p_{\pi^{0}}$ (top), energy asymmetry (middle), and $\cos~\theta_{\pi^{0}}$ (bottom) bins comparing Monte Carlo and data, each normalized to unit area. Error bars on data are statistical only. Monte Carlo error bars are estimated systematic errors from the yield extraction procedure {\bf only}, added in quadrature with statistical errors. Statistical errors alone are indicated by the innermost hashes on MC error bars.
\label{fig:NCpi0}}
\end{figure}

Understanding the rate and kinematics of $\pi^{0}$ production in neutrino neutral current interactions is critical for MiniBooNE, since these events are a major background for the $\nu_{\mu}\rightarrow\nu_{e}$ appearance search. The preliminary analysis of the mass spectrum and kinematics of $\pi^{0}$'s produced in neutral current interactions observed in MiniBooNE is presented here.

Fig.~\ref{fig:pi0_mass} shows the reconstructed invariant mass of beam triggers which satisfy the NC $\pi^0$ event selection criteria. As with the CCQE analysis, cosmic ray-induced activity during the beam trigger is eliminated by the method discussed in section \ref{sec:intro}. Further cuts on fiducial volume and energy of gammas from $\pi^0$ decay are applied to ensure a good reconstruction of the event. The fitted curves are Monte Carlo-based parameterizations, where the signal contribution arises from NC resonant and coherent single $\pi^{0}$ events, and the background contribution comes from all other events. The background shape is determined by Monte Carlo simulations. A background peak near $m_{\pi^{0}}$ is expected, since the background does contain some $\pi^{0}$'s produced in final state interactions and some $\pi^{0}$'s from multi-pion events. For approximately $1\times10^{20}$ protons on target, we extract $2425\pm107$ signal-like NC $\pi^{0}$ events in the fit to data in the preliminary analysis. The fitted mass peak agrees well with the nominal $\pi^{0}$ mass.

The number of neutral current $\pi^{0}$'s seen in data is also extracted in bins of calculated variables of interest ($p_{\pi^{0}}$, $cos~\theta_{\pi^{0}}$, $\frac{|E_{1}-E_{2}|}{E_{1}+E_{2}}$) to produce a distribution of each variable with the binned yields. The signal fraction in each bin is extracted via a fit to the invariant mass plot for events in that bin. The distributions for each variable are compared with expectations from Monte Carlo simulations, both distributions normalized to unit area.

Fig.~\ref{fig:NCpi0} (left) shows the $\pi^{0}$ momentum distribution for data and Monte Carlo. For all parts of this figure, the error bars shown on data are statistical only. Monte Carlo errors are statistical added in quadrature with estimated systematic errors stemming from the yield extraction procedure. The fraction of $\pi^{0}$'s reconstructed at high momentum drops off mainly due to the fall-off in the neutrino flux spectrum. 

The energy asymmetry of the $\gamma$'s from $\pi^{0}$ decay is shown in Fig.~\ref{fig:NCpi0} (middle). It falls off just below 0.9 because of the minimum energy requirement for each of the $\gamma$'s in the decay; events where one or both of the $\gamma$'s have less than 40 MeV are cut. The angular distribution of $\pi^{0}$'s relative to the beam direction is shown in Fig.~\ref{fig:NCpi0} (right). The distribution is sensitive to the production mechanism -- coherent $\pi^{0}$s are much more strongly forward-peaked than resonant $\pi^{0}$'s. This distribution may help us to determine how much coherent production contributes to the overall rate of NC $\pi^{0}$ production.

\subsection{Neutral Current Elastic Scattering Events}

\begin{figure}[ht]
\centerline{\epsfxsize=4.1in\epsfbox{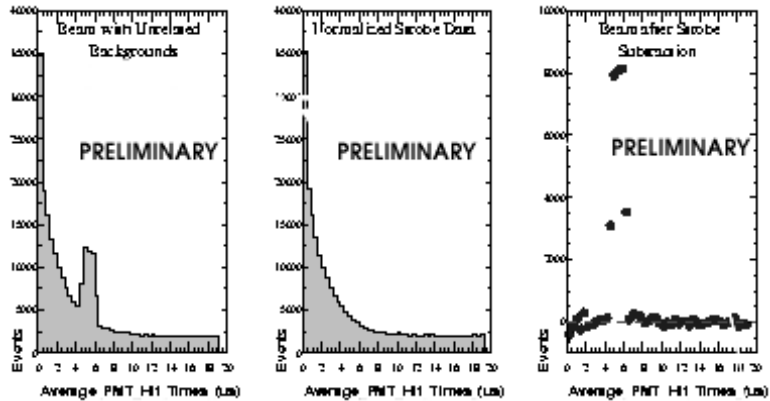}}
\caption{\label{fig:NCE_beamsubtr}Average event time strobe subtraction. In the left panel, the 1.6 $\mu$s wide peak located at 4.6 - 6.2 $\mu$s is from beam neutrinos. The center panel shows random strobe triggers. The right panel shows the beam window with random strobe triggers subtracted off.}
\end{figure}

\begin{figure}[ht]
\centerline{\epsfxsize=2.5in\epsfbox{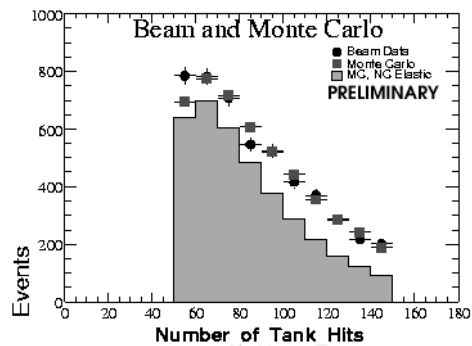}}
\caption{\label{fig:NCE_answer}Tank hits with strobe subtraction compared to relatively
normalized Monte Carlo for events with scintillation fraction $\geq$ 0.5.}
\end{figure}

MiniBooNE can also easily isolate $\nu_{\mu}$ neutral current elastic scattering events\cite{chris} with only a few simple cuts. The primary background to NCE scattering events is cosmic muons, and the subsequent Michel electrons from the muon decay.  Random strobe triggers, which contain the beam-unrelated backgrounds from cosmics and environmental radioactivity, are used to make a ``beam-off'' subtraction. The method is equivalent to that described in section \ref{sec:intro}, but the cuts are optimized for very low energy events. This is shown in Fig.~\ref{fig:NCE_beamsubtr}. In the left panel, note the 1.6-$\mu$s-wide peak due to beam neutrino interactions. The falling exponential is caused by the decay of cosmic muons. Once the subtraction is performed, a flat distribution, except for the expected beam spike, remains. 

Neutral current elastic interactions should produce relatively low energy nucleons, and therefore a low number of tank hits. They are also expected to leave a relatively high amount of scintillation light in the detector, since only nuclear fragments are detected. Fig.~\ref{fig:NCE_answer} shows the strobe-subtracted distributions of tank hits for events with scintillation fraction $>0.5$.  The Monte Carlo is relatively normalized to the data for shape comparison. The histogram shows the relatively high fraction of neutral current events expected in this sample.

\section{Expectations for the Future}

MiniBooNE has been running smoothly for more than a year now. As of February, 2004, the experiment has collected over 200,000 contained neutrino candidates with the $1.8 \times 10^{20}$ protons on target received. We expect to have $5 \times 10^{20}$ protons on target by June, 2005. The three non-oscillation analyses described in this paper, CCQE, NC $\pi^0$, and NCE, are well underway. The main items remaining to be studied and understood for these are the systematic errors associated with the $\nu_{\mu}$ flux, $\nu_{\mu}$ cross sections, and the properties of light production and transmission in the MiniBooNE detector. These analyses are necessary for determining the two oscillation results, $\nu_{\mu}$ disappearance and $\nu_{e}$ appearance, which are the main goals of this experiment.

\end{document}